\RequirePackage{fix-cm}
\documentclass{svjour3}                     
\smartqed  
\usepackage{graphicx}
\usepackage{epstopdf}
\usepackage{epsfig}
\usepackage{natbib}
\usepackage{color}
\usepackage{amsmath}

 \journalname{Experiments in Fluids}
\begin{document}

\title{Optical-flow-based background-oriented schlieren technique
for measuring a laser-induced underwater shock wave
}
\subtitle{}


\author{Keisuke Hayasaka$^{1}$, Yoshiyuki Tagawa$^{1*}$, \\Tianshu Liu$^{2,3}$, Masaharu Kameda$^{1,3}$}

\authorrunning{Keisuke Hayasaka, Yoshiyuki Tagawa, Tianshu Liu, Masaharu Kameda} 

\institute{
1: Department of Mechanical Systems Engineering, Tokyo University of Agriculture and Technology, Japan  \\
2: Department of Mechanical and Aerospace Engineering, Western Michigan University, Kalamazoo, MI 49008, USA \\
3: Institute of Global Innovation Research, Tokyo University of Agriculture and Technology \\\\
          $^{*}$Correspondent author: Y. Tagawa \at
              \email{tagawayo@cc.tuat.ac.jp}\\
              Tel.: +8142-388-7407\\
              Fax: +8142-388-7407\\
}

\date{Received: date / Accepted: date}

\maketitle

\begin{abstract}
The background-oriented schlieren (BOS) technique with the physics-based optical flow method (OF-BOS) is developed for measuring the pressure field of a laser-induced underwater shock wave.  
Compared to BOS with the conventional cross-correlation method in PIV (called PIV-BOS), by using the OF-BOS, the displacement field generated by the small density gradient in water can be obtained at the spatial resolution of one vector per pixel.  
The corresponding density and pressure fields can be further extracted.  
It is particularly demonstrated that the sufficiently high spatial resolution of the extracted displacement vector field is required in the tomographic reconstruction to correctly infer the pressure field of the spherical underwater shock wave.  
The capability of the OF-BOS is critically evaluated based on synchronized hydrophone measurements. 
Special emphasis is placed on direct comparison between the OF-BOS with the PIV-BOS.
\keywords{Background-oriented schlieren (BOS) \and Optical flow \and Cross-correlation \and Particle image velocimetry (PIV), Underwater shock wave}
\end{abstract}

\section{Introduction}
\label{intro}
Non-contact measurements of underwater shock waves are crucial for understanding some important phenomena related to non-invasive medical treatments (\cite{klaseboer2007interaction, tagawa2012highly, tagawa2013needle}).  
The back-ground-oriented schlieren (BOS) technique can provide global non-contact diagnostics for measuring the displacement field induced by the density change in fluid, which is used to infer the density and pressure fields.  
The BOS technique has been used in various measurements particularly in compressible air flows and gas-mixing flows (\citet{raffel2000applicability, meier2002computerized, venkatakrishnan2004density, murphy2011piv}).  
A review of the BOS technique is given by \citet{raffel2015background}.  
The BOS technique is an extension of the classical schlieren technique traditionally used in high-speed wind tunnel for aerodynamics measurements.  
Due to the significant advance of digital cameras and image processing, the experimental setup of the BOS system is much simpler than its classical counterpart with complex arrangements of optics.  
In general, a background pattern plate placed behind a measurement domain of fluid is imaged by a camera in the cases with and without the fluid density gradient.  
The background pattern image is disturbed by the density gradient that deflects the light rays radiated from the background pattern plate.  
The disturbed image appears shifted relative to the undisturbed image taken in the case with the homogenous fluid density.  
The displacement field in the disturbed image is related to the path integral of the density gradient.  
When the displacement field is measured, the density and pressure fields could be inferred.  
To determine the displacement field from the undisturbed and disturbed images, the cross-correlation method in particle image velocimetry (PIV) can be applied. 

However, the application of the BOS technique in liquid is more difficult compared to its application in gas flows since the light-ray deflection due to the density change in liquids is much smaller.  
In particular, the measurement of an underwater shock wave by using the BOS technique is scanty.
Recently, using the BOS technique with the cross-correlation method, we obtained the displacement field induced by the local density gradient of the underwater shock wave (\citet{yamamoto2015application}).
Based on these results, we tried to further reconstruct the pressure field from the displacement field. 
However, it is found that the pressure field could not be correctly reconstructed due to the insufficient number of the displacement vectors extracted by the typical correlation algorithm in PIV (\citet{hayasakameasurement}).  
In order to overcome this problem, we propose a high-resolution BOS technique by incorporating the optical flow method, which is simply referred to as OF-BOS.  
\citet{atcheson2009evaluation} suggested that the typical optical flow methods in computer vision, such as the Horn-Schunck method and the Locas-Kanade method, could be used for the BOS technique.  
They found that these optical flow methods particularly the Horn-Schunck method performed better than the correlation method in terms of the accuracy and the spatial resolution.  

In this paper, we adopt the physics-based optical flow method in BOS, which is developed by \citet{liu2008fluid} for various flow visualizations.  
This method could achieve the spatial resolution of one vector per pixel and the better accuracy (\citet{liu2008fluid}), which is important for the application of the BOS technique in water. 

Following this instruction, we will describe the optical-flow-based BOS technique, tomographic reconstruction of the field of the divergence of the displacement vector, and determination of the density field by solving the Poisson's equation.  
Then, the experimental setup will be described for BOS measurements of the laser-induced shock wave.  
Next, the results will be discussed, focusing on the effects of the spatial resolution of the extracted displacement field on reconstruction of the density and pressure fields.  
In particular, we will quantitatively compare the pressure distributions obtained from both the OF-BOS and the PIV-BOS with hydrophone data.


\section{Optical-flow-based BOS technique}
\label{sec:OF-BOS}


\subsection{BOS}
\label{sec:BOS}

\begin{figure}[b]
\begin{center}
\includegraphics[width=0.8\textwidth]{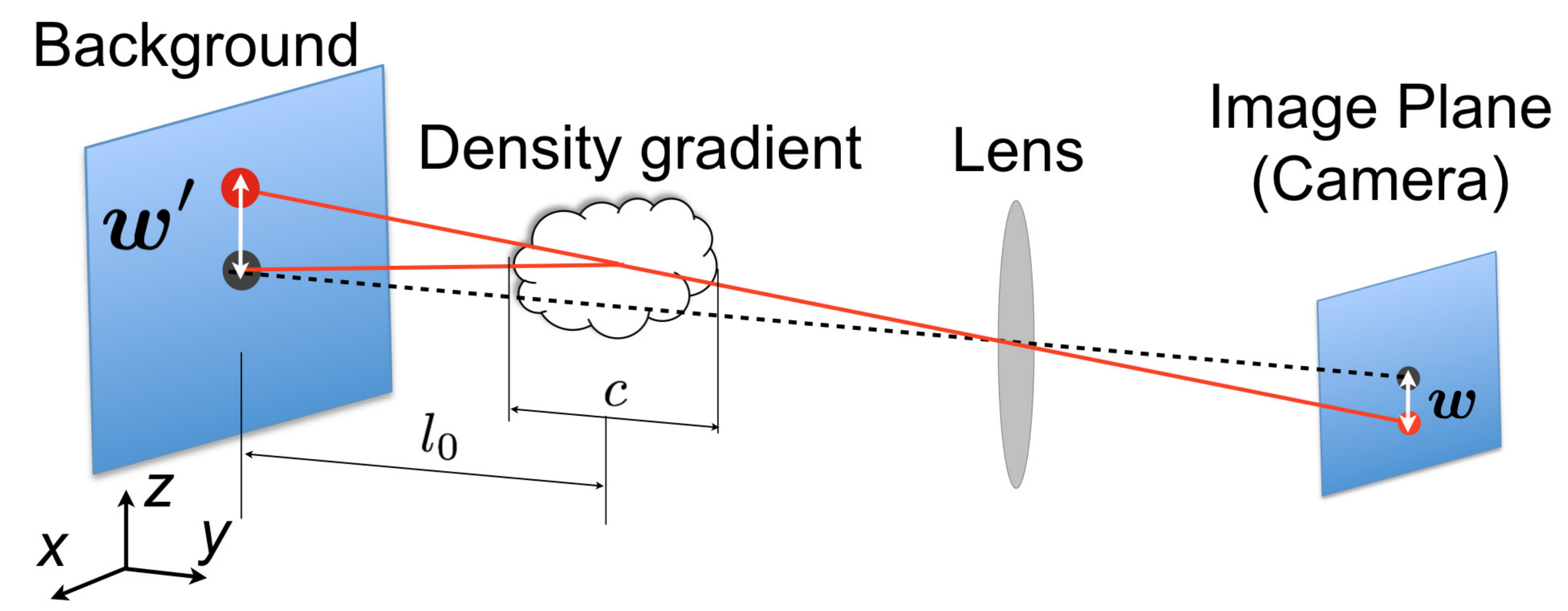}
\caption{Illustration of the principle of the BOS technique.}
\label{Fig1}
\end{center}       
\end{figure}

\begin{figure*}[h]
\includegraphics[width=1.00\textwidth]{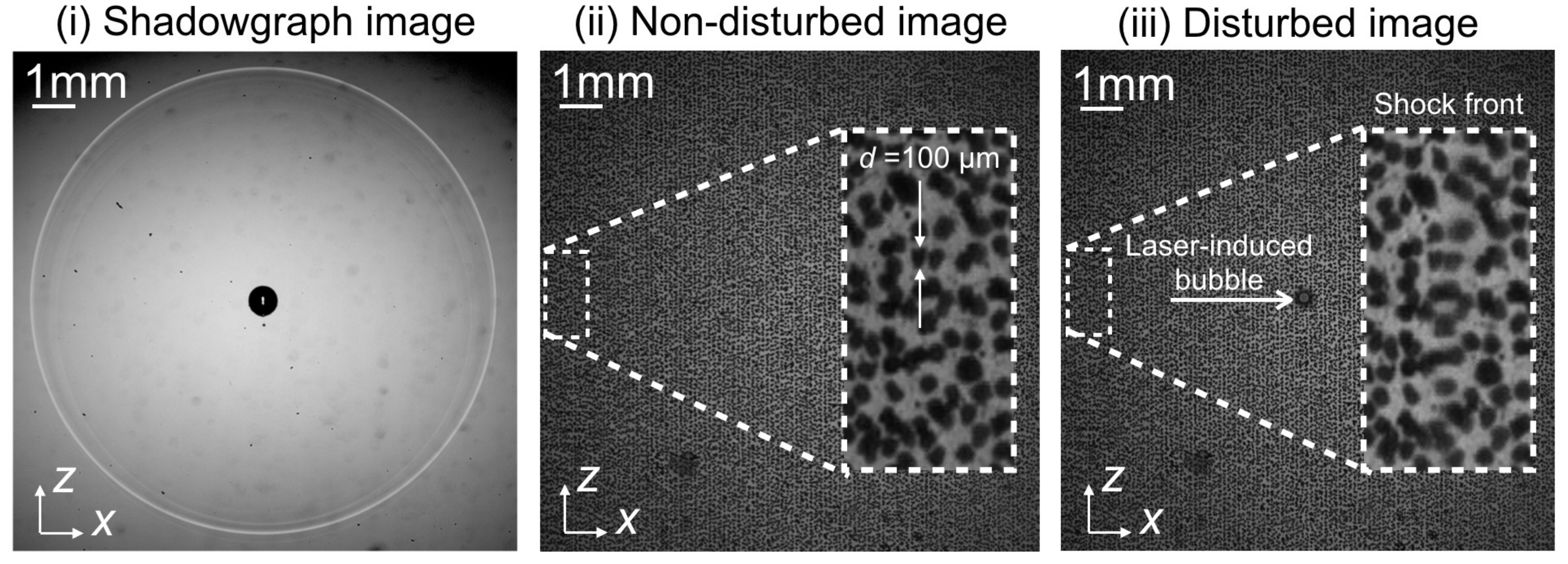}
\caption{Typical images obtained the experimental setup in this work: (i) A shadowgraph snapshot of the laser induced shock wave, (ii) non-disturbed image in BOS, and (iii) disturbed image in BOS.}
\label{Fig2}       
\end{figure*}

The basic equations in the BOS technique are briefly recapitulated for convenience of reading.  
The working principle of the BOS technique is illustrated in Fig. \ref{Fig1} (\citet[]{venkatakrishnan2004density, yamamoto2015application}).  
A typical BOS system consists of a camera, a background pattern plate (such as a random dot pattern plate), a light source, and a computer for image/data processing.  
For simplicity, it is assumed that the background reference plane $(x,z)$ is parallel to the image plane.
Thus, the image coordinates are equal to $\lambda$$(x,z)$, where $\lambda$ is a proportional constant in the orthographical projection.  
The light ray radiated from the background pattern plate is deflected from the original straight path through a fluid domain with the density gradient due to optical refraction, and therefore the image of the background pattern has an apparent displacement (shift) field.  

The displacement on the image plane is proportional to the path integral of the small density gradient through a fluid domain.  
For BOS setting, \citet{van2014density} gave the following relation for the in-plane displacement vector ${\mbox{\boldmath $w$}}^{\prime}$ in the background plate

\begin{equation}
\label{eq1}
{\mbox{\boldmath $w$}}^{\prime} =\frac { 1 }{ 2 } c(c+2{ l }_{ 0 })\frac { 1 }{ { n }_{ 0 } } \nabla n,
\end{equation}
\noindent
where $c$ (100 $\mu$m in this case) is the thickness of the density gradient domain, $l_{0}$ (8 mm in this case) is the distance from the density gradient domain to the background plate, $n_{0}$ is the refractive index of water at the condition without a shock wave, $n$ is the refractive index of water in the test condition, and $\nabla$ is the gradient operator on the coordinate plane $(x,z)$.  
The relation between the refractive index and the fluid density is given by the Gladstone-Dale equation (\citet[]{merzkirch2012flow, raffel2015background}) 

\begin{equation}
\label{eq2}
n=K\rho +1,
\end{equation}
\noindent
where $K$ is the Gladstone-Dale constant (3.34$\times$10$^{-4}$ m$^{3}$/kg for water), and $\rho$ is the density of the fluid.  
Substitution of Eq. (2) into Eq. (1) yields

\begin{equation}
\label{eq3}
{\mbox{\boldmath $w$}}^{ \prime  }=\frac { 1 }{ 2 } c(c+2{ l }_{ 0 })\frac { K }{ 1+K{ \rho  }_{ 0 } } \nabla \rho ,
\end{equation}
\noindent
where $\rho_{0}$ is the fluid density under hydrostatic pressure.  
It is emphasized that the displacement vector ${\mbox{\boldmath $w$}}^{\prime}$ obtained by the BOS technique is the path-integrated (or projected) quantity across the measurement domain with the density gradient.  
In general, to extract the 3D field from the path-integrated quantity requires tomographic reconstruction from data at multiple viewing directions.  

Further, applying the dot product of $\nabla$ to Eq. (3), we have the Poisson's equation for the density 

\begin{equation}
\label{eq4}
{ \nabla  }^{ 2 }\rho =S,
\end{equation}
\noindent
where the source term $S$ is proportional to the divergence of the displacement vector ${\mbox{\boldmath $w$}}^{\prime}$ by

\begin{equation}
\label{eq5}
S=\frac { 2(1+K{ \rho  }_{ 0 }) }{ cK(c+2{ l }_{ 0 }) } \nabla \cdot {\mbox{\boldmath $w$}}^{\prime}.
\end{equation}

In principle, when the displacement vector ${\mbox{\boldmath $w$}}^{\prime}$ is measured, the density field can be determined by solving Eq. (4).  
The Gauss-Seidel method is used to solve Eq. (4) numerically here.  
Then, the pressure field can be further determined by applying the Tait equation

\begin{equation}
\label{eq6}
\frac { p+B }{ { p }_{ 0 }+B } ={ \left( \frac { \rho  }{ { \rho  }_{ 0 } }  \right)  }^{ \alpha  },
\end{equation}
\noindent
where $p_{0}$ is the hydrostatic pressure, $B$ is a constant of 314 MPa, and the exponent $\alpha$ is 7 (\citet[]{brujan2010cavitation, yamamoto2015application}).  
In this study, we measure a spherical shock wave in water with the peak overpressure of about 1 MPa.  
In this case, the density change due to the overpressure is less than 1 kg/m$^{3}$.  
This means that the change in the refractive index is very small [O(10$^{-4}$)].  

The BOS system acquires an undisturbed background pattern image and the corresponding disturbed image.  
Figure \ref{Fig2} shows the typical images obtained in the experimental setup in this work.  
Figure \ref{Fig2}(i) shows a shadowgraph snapshot of a shock wave propagating spherically around a laser-induced bubble, where the high-pressure region is expected at a shock front.  
Figure \ref{Fig2}(ii) and \ref{Fig2}(iii) show an undisturbed background reference image and the corresponding disturbed image, respectively.  
The zoomed-in image of the particle pattern is shown as well.  
Without image processing, it is difficult to see directly the shock wave in Fig. \ref{Fig2}(iii) since the change of the refractive index of water is so small.


\subsection{Optical flow method}
\label{sec:optical flow}

\begin{figure}[b]
\includegraphics[width=0.8\textwidth]{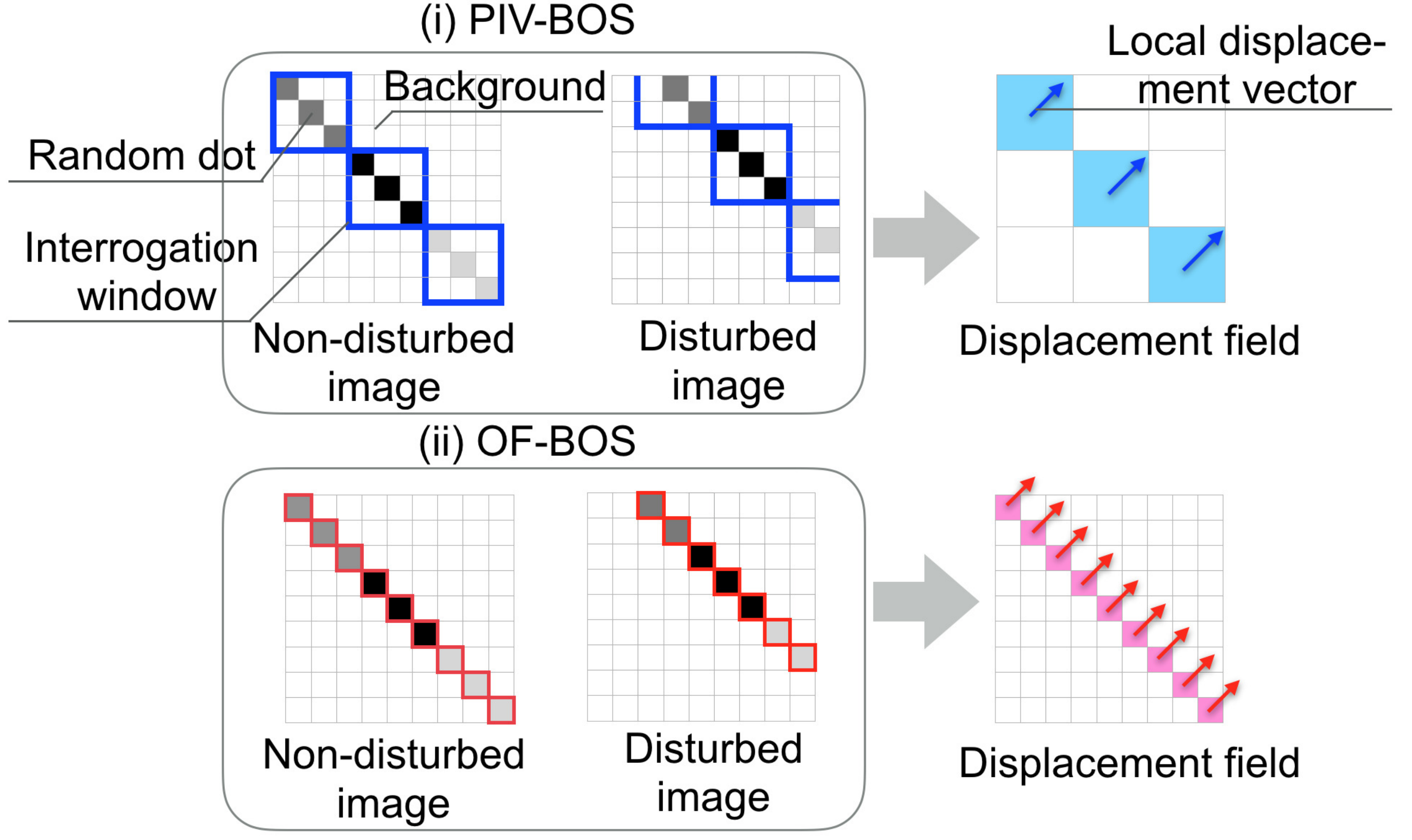}
\caption{Illustration of measurements of displacement vectors. (i) PIV-BOS (ii) OF-BOS.}
\label{Fig3}       
\end{figure}

The key element in the BOS technique is to determine the displacement vector field in the undisturbed background reference image and the corresponding disturbed image.  
This computation is usually carried out by using the cross-correlation method in PIV.  
An effort has been made to adopt the classical optical flow methods in computer vision science in BOS (\citet{atcheson2009evaluation}).  
The optical flow method can be further developed based on rational physical foundations for various flow visualizations (\citet{liu2008fluid}). 
The optical flow is described by the generic equation in the image plane, i.e.,

\begin{equation}
\label{eq7}
\frac { \partial g }{ \partial t } +\nabla \cdot \left( g \bf{u}\right) =f,
\end{equation}
\noindent
where $\bf{u}$ is the velocity in the image plane referred to as the optical flow, $g$ is the normalized image intensity, $\nabla$ is the spatial gradient in the image plane, and $f$ is a term related to the diffusion and boundary fluxes which are negligibly small in most cases.  
In velocimetry, the optical flow in Eq. (7) has a clear physical meaning, that is, the optical flow is proportional to the light-ray-path-averaged velocity of fluid or particles in flows.  
We call this method as the physics-based optical flow technique.  
In a special case where $\nabla$$\cdot$$\bf{u}$ = 0 and $f$ = 0, Eq. (7) is reduced to the Horn-Schunck brightness constraint equation which is the foundation of the classical optical flow method developed by \citet{horn1981determining}.  
It is noted that in a general case the optical flow is not divergence-free and $\nabla$$\cdot$$\bf{u}$ = 0 is not physically true.  
For BOS applications, the difference $\partial g$/$\partial t$ $\approx$ $\Delta g$/$\Delta t$ is used in Eq. (7), where $\Delta g$ = $g$ - $g_{ref}$ is the difference between the disturbed image and the undisturbed reference image, and the nominal time interval is unitary ($\Delta t$ = 1).  
Therefore, the optical flow in Eq. (7) is interpreted as the displacement vector field in the image plane (i.e.  $\bf{u}$ = ${\mbox{\boldmath $w$}}$)  that is generated by the deflected light ray through the density gradient domain.  
The displacement vector in the background plane is related to that in the image plane by ${\mbox{\boldmath $w$}}$ = $\lambda$${\mbox{\boldmath $w$}}^{\prime}$. 

To determine the displacement vector field in the image plane, a variational formulation with a smoothness constraint is typically used (\citet{liu2008fluid}).  
By minimizing the functional, the Euler-Lagrange equation is given for the optical flow.  
The standard finite difference method is used to solve the Euler-Lagrange equation with the Neumann condition on the image domain boundary for the optical flow.  
The optical flow algorithm used in this work has the routines: the Horn-Schunck estimator for an initial solution (\citet{horn1981determining}) and the Liu-Shen estimator for a refined solution of Eq. (7) (\citet{liu2008fluid}).  
The relevant parameters in pre-processing and optical flow computation should be suitably selected.  
The main parameters are the Lagrange multipliers for the Horn-Schunck and Liu-Shen estimators.  
Other parameters are the number of iterations in successive improvement of optical flow computation by using a coarse-to-fine iterative scheme, and the sizes of the Gaussian filters for correcting the effect of a local illumination intensity change and removing small random noise in images.  
A mathematical analysis of the physics-based optical flow and an iterative numerical algorithm are given by \citet{wang2015analysis}.  
Quantitative comparison between the optical flow and cross-correlation methods for PIV images has been evaluated by \citet{liu2015comp}.  
The variational optical flow method based on Eq. (7) is a differential approach that can achieve the theoretical spatial resolution of one vector per pixel.  
Figure \ref{Fig3} illustrates the difference between the optical flow method as a differential approach and the cross-correlation method as a region-based integral approach.  
The optical flow method is particularly suitable to detect small displacements in narrow regions (such as shock wave).


\subsection{Tomographic reconstruction on a spherical surface}
\label{sec:tomographic}

\begin{figure}[b]
\begin{center}
\includegraphics[width=0.5\textwidth]{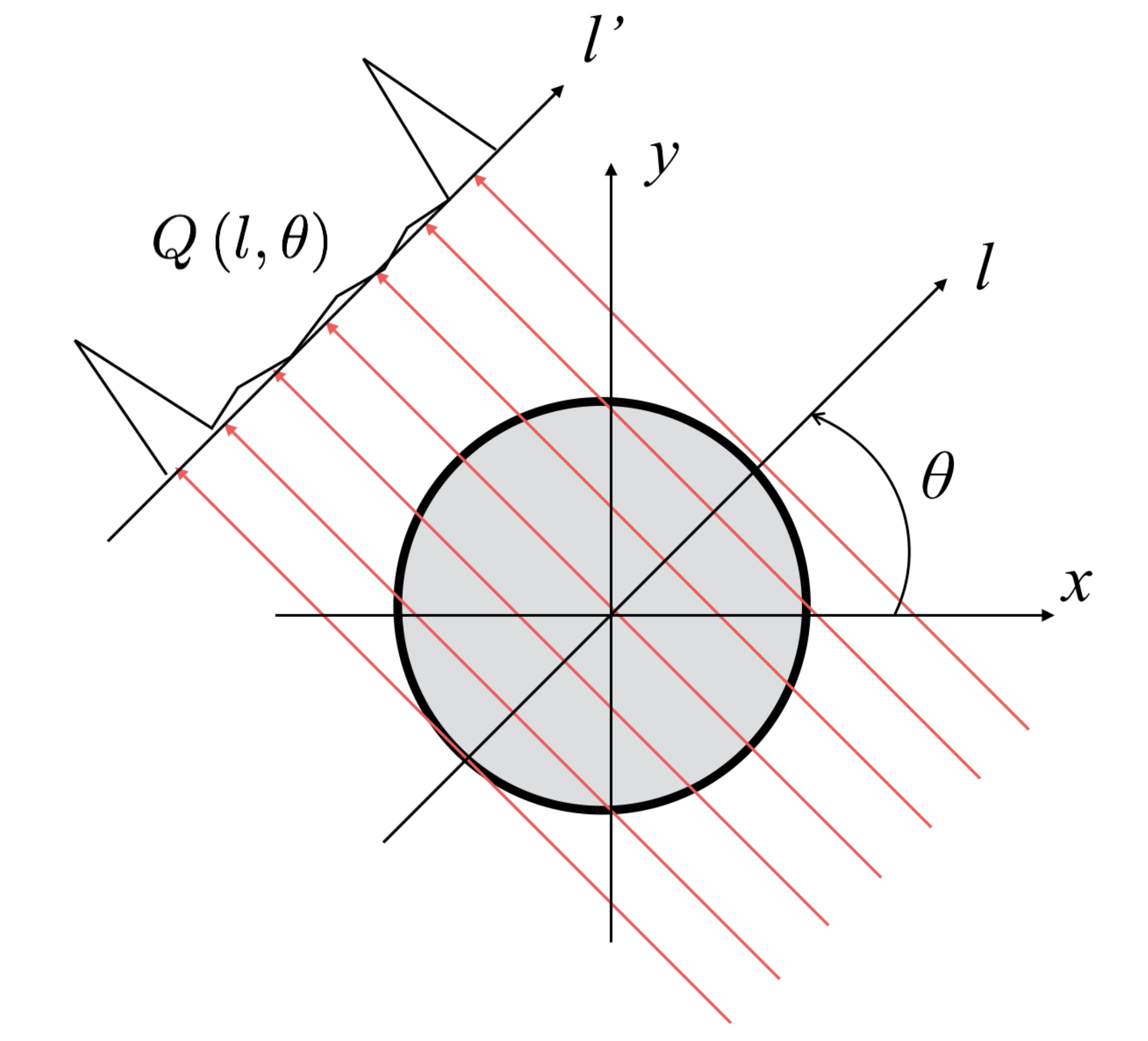}
\caption{Projections of light rays through a domain on the image plane, where $l'$ is an axis on the image plane.}
\label{Fig4}  
\end{center}     
\end{figure}

\begin{figure*}[h]
\begin{center}
\includegraphics[width=1\textwidth]{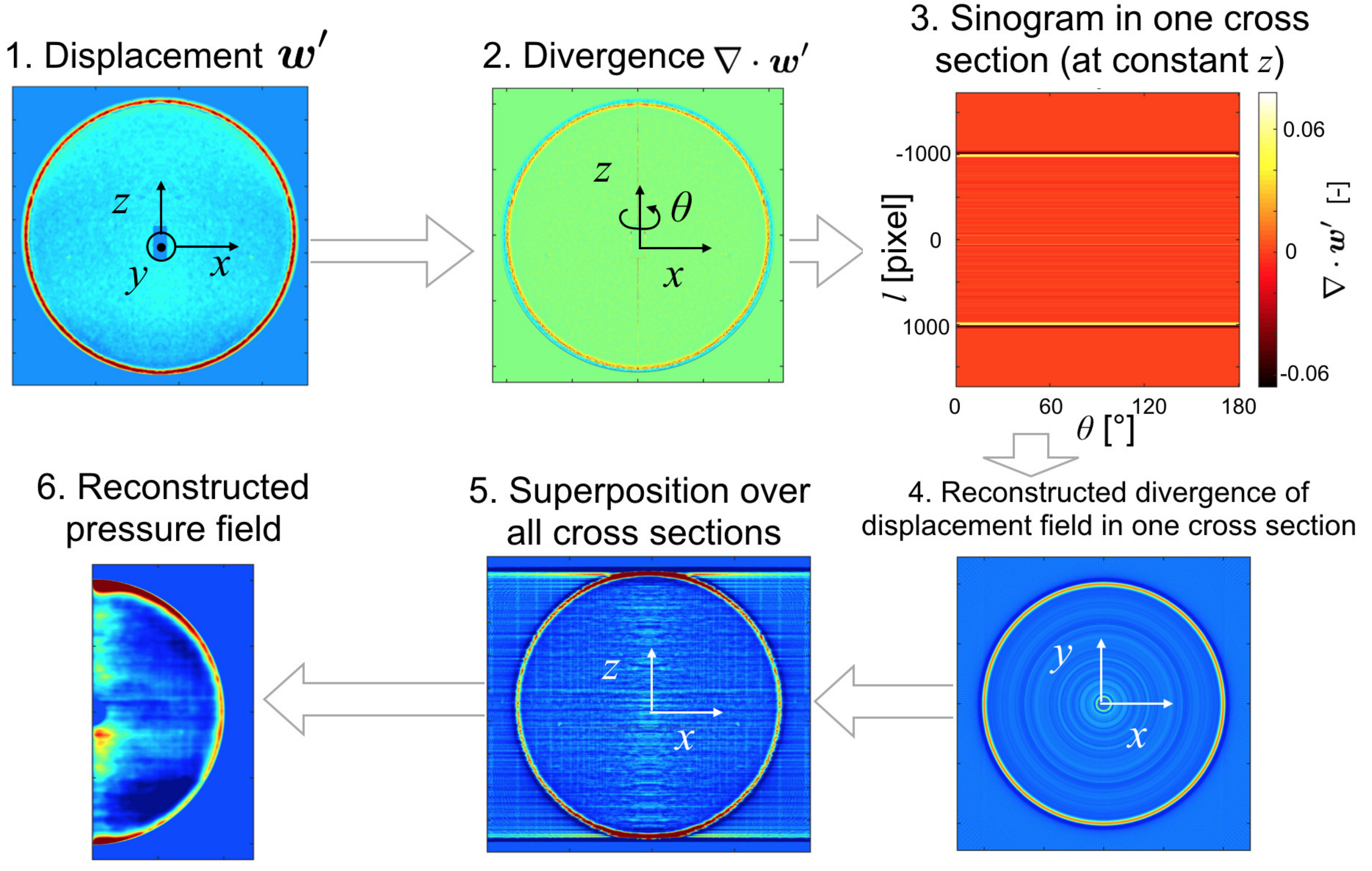}
\caption{Tomographic reconstruction procedures for a spherical shock wave.}
\label{Fig5}   
\end{center}    
\end{figure*}

\begin{figure}[b]
\includegraphics[width=0.8\textwidth]{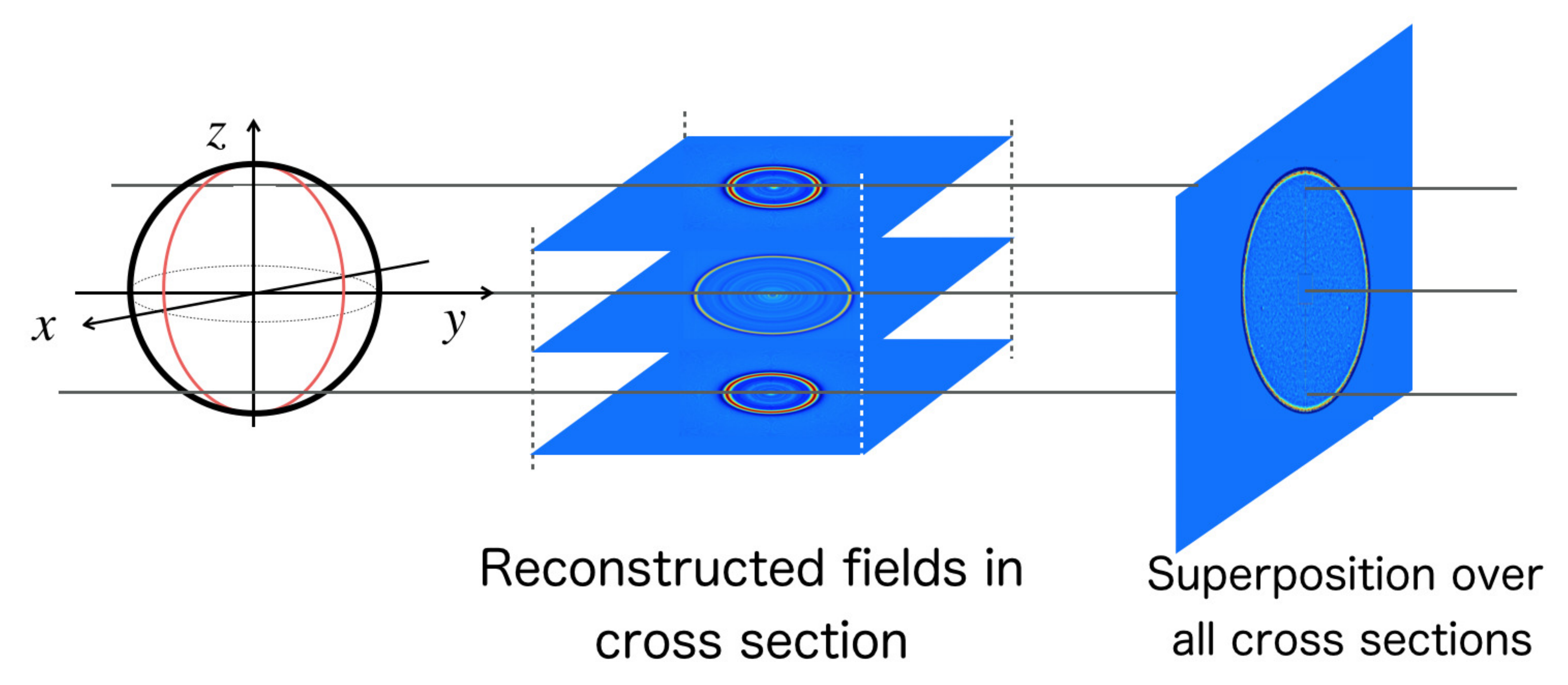}
\caption{Superposition of reconstructed fields at all cross sections in the $z$-coordinate.}
\label{Fig6}       
\end{figure}

The tomographic reconstruction technique is used to reconstruct the relevant physical quantity (such as the fluid density or the divergence of the displacement vector) on a spherical surface of a propagating shock wave from the path-integrated or projected quantity obtained from the BOS measurements.  
In volumetric flow visualizations, an image can be modeled as a set of 1D projections of a field of a quantity with the given projection angles through a 2D domain as illustrated in Fig. \ref{Fig4}.  
The basic tomographic problem is how to reconstruct the 2D field from these projections.  
The projection of the quantity $q(x,y)$ can be expressed as an integral along a ray, i.e.,

\begin{align}
\begin{split}
R\left( q \right) & \equiv Q\left( l,\theta  \right)\\
& = \int _{ -\infty  }^{ \infty  }{ \int _{ -\infty  }^{ \infty  }{ q\left( x,y \right) \delta \left( x\cos { \theta  } +y\sin { \theta  } -l \right) dxdy }  }, 
\end{split}
\end{align} 
\noindent
where $l=x\cos \theta+y\sin \theta$ is the coordinate along a projection line, $\theta$ is the angle defining the projection lines, $R(q)$ denotes the Radon transform, and $\delta$ denotes the Dirac-delta function. 
An inversion of the Radon transform is sought for $q(x,y)$. 
This tomographic problem has been studied in 3D flow measurements (\citet[]{feng2002visualization, venkatakrishnan2004density}).  
We introduce the Fourier transform

\begin{equation}
\label{eq9}
S_{\theta}( \xi ) =\int _{ -\infty  }^{ \infty  }{ Q\left( l,\theta  \right) { e }^{ -j2 \pi \xi l }dl }.
\end{equation}
\noindent
Then, the solution of the integral equation Eq. (8) can be formally given by

\begin{equation}
\label{eq10}
q\left( x,y \right) =\int _{ 0 }^{ \pi  }{ \int _{ -\infty  }^{ \infty  }{ S_{\theta}\left( \xi \right)  { e }^{ j2 \pi \xi l } \left| \xi  \right|d\xi d\theta  }  }.
\end{equation}
\noindent
Clearly, the tomographic reconstruction of $q(x,y)$ requires many projections in $\theta$ $\in$ $[0,\pi]$.  
In actual computation, the inverse Radon transform in Matlab is used in this work.  
In this process, we generate the sinogram digitally.  
Although the Abel transform is simpler for the reconstruction of an axial symmetrical field, we do not utilize it because of its high sensitivity to noise (\citet{venkatakrishnan2004density}).  

In the BOS measurement of a spherical shock wave, the following procedures are proposed as illustrated in Fig. \ref{Fig5}.

\begin{enumerate}
  \item Image registration based on the affine transformation is applied to the reference and disturbed images to correct any global misalignment between them caused by the possible movement of the camera and other factors during measurements.
  \item The projected quantity is selected as the divergence of the projected displacement vector, i.e., $Q(l,\theta)$ = $\nabla \cdot {\mbox{\boldmath $w$}}^{\prime}$ from the BOS measurement since it is the source term of the Poisson's equation for the fluid density.  
  The quantity $\nabla \cdot {\mbox{\boldmath $w$}}^{\prime}$ is mapped onto the image plane based on an assumption of the axial symmetrical structure of the laser-induced shock wave.  
  This is illustrated in Step (1-2) in Fig. 5. 
  \item The Radon transform $Q(l,\theta)$ = $\nabla \cdot {\mbox{\boldmath $w$}}^{\prime}$  is given as a sinogram at section at a given $z$-coordinate, as illustrated in Step (2-3).  
  In the sinogram, we generate 180 projected data in $\theta$ $\in$ $[0,\pi]$.
  \item The distribution of $q(x,y)$ at that section is reconstructed by using the inverse Radon transform (symbolically expressed as $q(x,y)$ = $R^{-1}$$(Q)$), as illustrated in Step (3-4).  
  In Matlab computation, we apply the spline interpolation to the back projection method and calculate the inverse Radon transform without filtering.
  \item The field of $\nabla \cdot {\mbox{\boldmath $w$}}^{\prime}$ on the spherical shock wave is obtained by superposition of the reconstructed fields over a set of cross sections in the $z$-coordinate, as illustrated in Step (4-5).  
  This superposition procedure for a spherical shock wave is further illustrated in Fig. \ref{Fig6}.
  \item The field of the density is obtained by solving the axisymmetric Poisson's equation for a given source term on the plane of symmetry and the corresponding pressure field is calculated by using Eq. (6), as shown in Step (5-6).
\end{enumerate}

It is necessary to comment the selection of the projected quantity for tomographic reconstruction.  
In the BOS measurements by \citet{venkatakrishnan2004density}, the projected fluid density was obtained first by solving the Poisson's equation, and then the local density was reconstructed by using the tomographic technique.  
In contrast, the present procedures conduct the tomographic reconstruction before solving the Poisson's equation.  
The projected quantity is the divergence of the projected displacement vector, i.e., $Q(l,\theta)$ = $\nabla \cdot {\mbox{\boldmath $w$}}^{\prime}$, and thus the divergence of the local displacement vector is first reconstructed by using the tomographic technique.  
Next, the local density is calculated by solving the Poisson's equation.  
In principle, since the Poisson's equation is valid for both the local and path-integrated (projected) quantities, the approach used by \citet{venkatakrishnan2004density} should be equivalent to the present approach.  
Nevertheless, the rationale behind the present arrangement is to avoid the propagation of the error in solving the Poisson's equation into tomographic reconstruction that is more sensitive to the errors.


\section{Experimental setup}
\label{sec:Experimental setup}

\begin{figure}[b]
\includegraphics[width=0.8\textwidth]{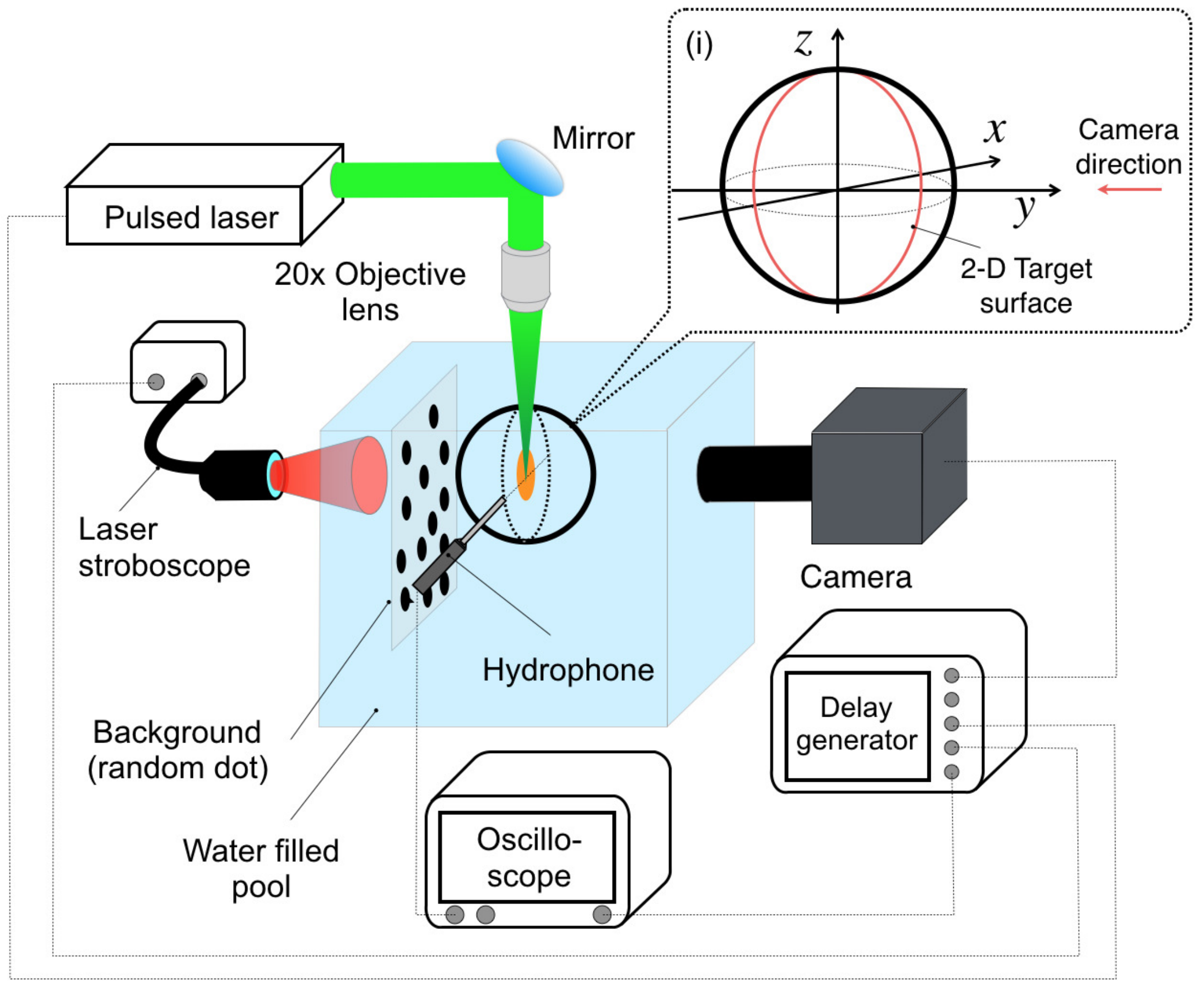}
\caption{Experimental setup for BOS measurements of the laser-induced shock wave.}
\label{Fig7}       
\end{figure}

Figure \ref{Fig7} is a schematic of the experimental setup of the BOS measurements of a shock wave generated by a pulsed laser.  
A laser pulse with the wavelength of 532 nm and pulse width of 6 ns (Nd:YAG laser Nano S PIV, Litron Lasers Ltd.) was focused through a 20$\times$microscope objective lens to a point inside a distilled-water-filled glass tank (450$\times$300$\times$300 mm$^3$).  
The underwater shock wave was generated at a laser-focused point, propagating spherically.  
The background random-dot-pattern plate was placed behind the laser-focused point inside the tank.  
The shock wave was recorded by using a camera with 2048$\times$2048 pixels (FASTCAM Mini WX50, Photron Ltd.).  
The spatial resolution was set at 5.18 $\mu$m per pixel.  
The light source utilized for illuminating the background pattern was a laser stroboscope (SI-LUX 640, Specialized Imaging Ltd.) with the pulse width of 20 ns.  
The pulsed laser, camera, and light source were synchronized by using a delay function generator (Model 575, BNC Co.).  
We utilized a PVDF pressure sensor (i.e. hydrophone) (Muller-Platte Needle Prove, Mueller Instruments) to validate the results obtained by using the OF-BOS and the PIV-BOS. 
The pressure sensor was set toward a center of the laser-induced shock.  
The distance from the center to the hydrophone was 5.0$\pm$0.1 mm.  
In the PIV-BOS, the open-source Matlab PIV code called PIVlab is used to determine the displacement vectors, which is described by \citet{thielicke2014pivlab}. 


\section{Results}
\label{sec:Reslts}


\subsection{Typical case}
\label{sec:Typical case}

\begin{figure}[b]
\includegraphics[width=0.8\textwidth]{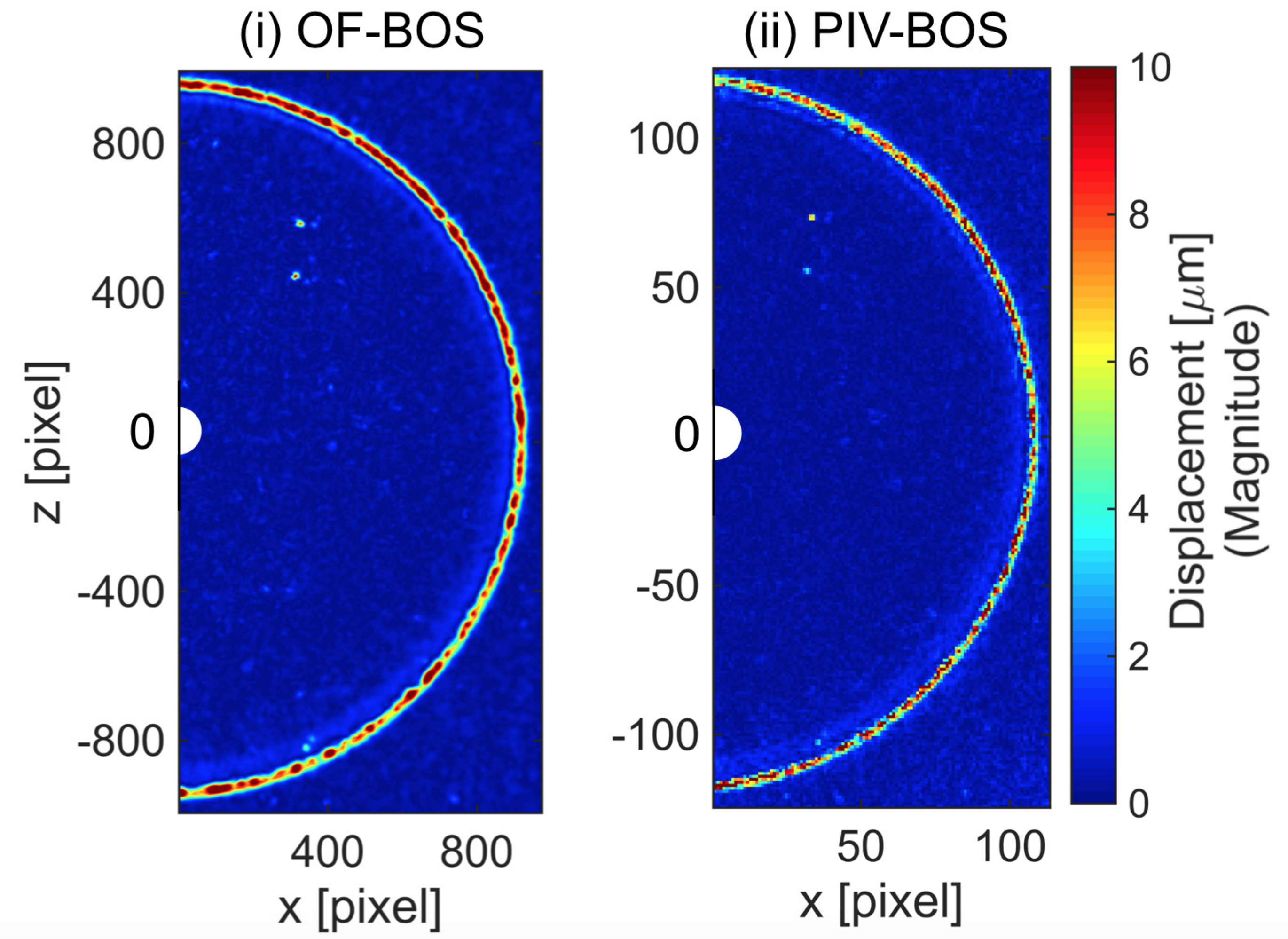}
\caption{The displacement magnitude fields in the half domain: (i) OF-BOS, and (ii) PIV-BOS.}
\label{Fig8}       
\end{figure}

\begin{figure}[b]
\includegraphics[width=0.8\textwidth]{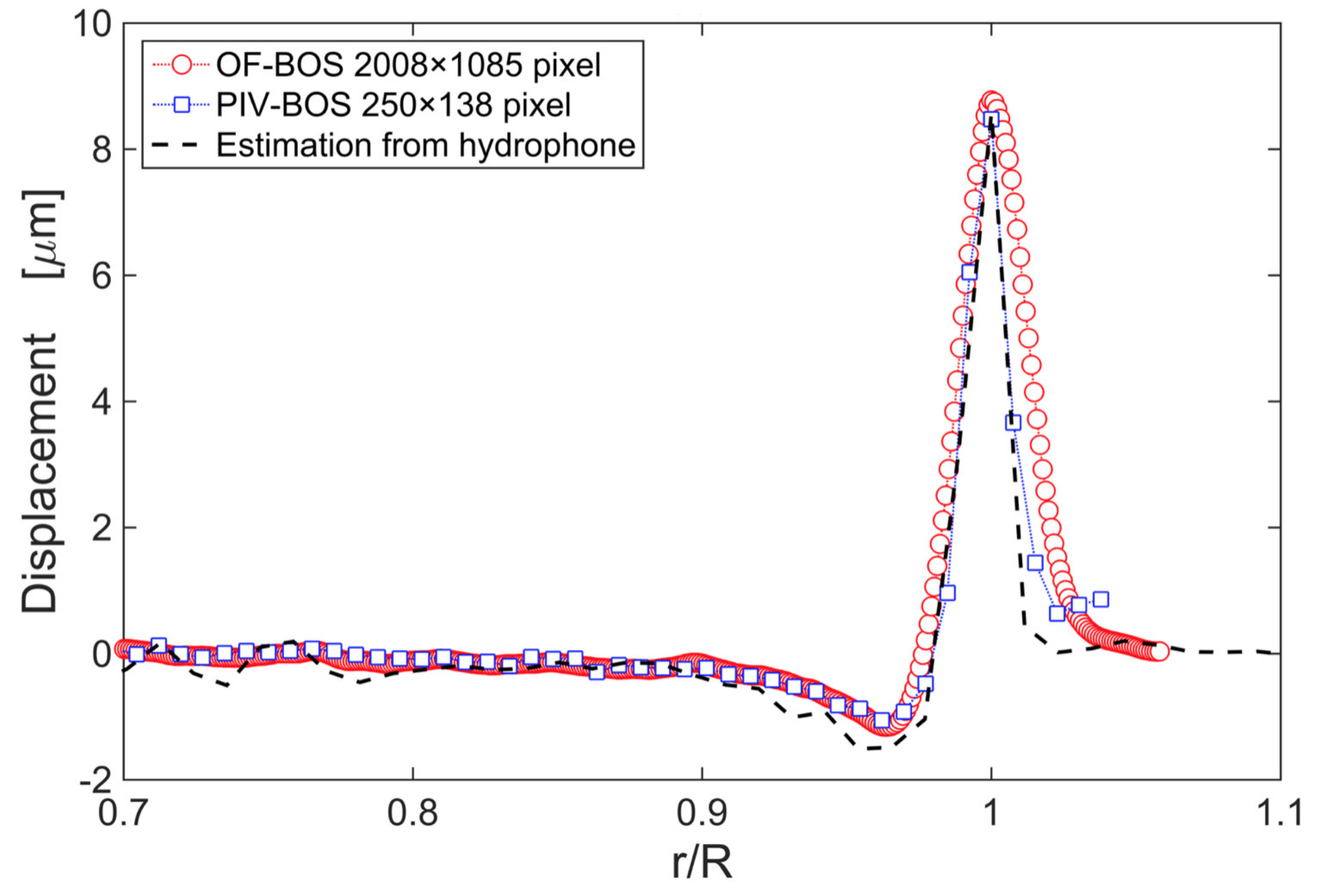}
\caption{The profiles of the measured displacement along a ray aligned with the hydrophone, where $R$ is the radius of the shock wave, where the averaging operation over 50 lines for the OF-BOS and 20 lines for the PIV-BOS in the $z$-direction is applied to reduce the noise.}
\label{Fig9}       
\end{figure}

\begin{figure}[b]
\includegraphics[width=0.8\textwidth]{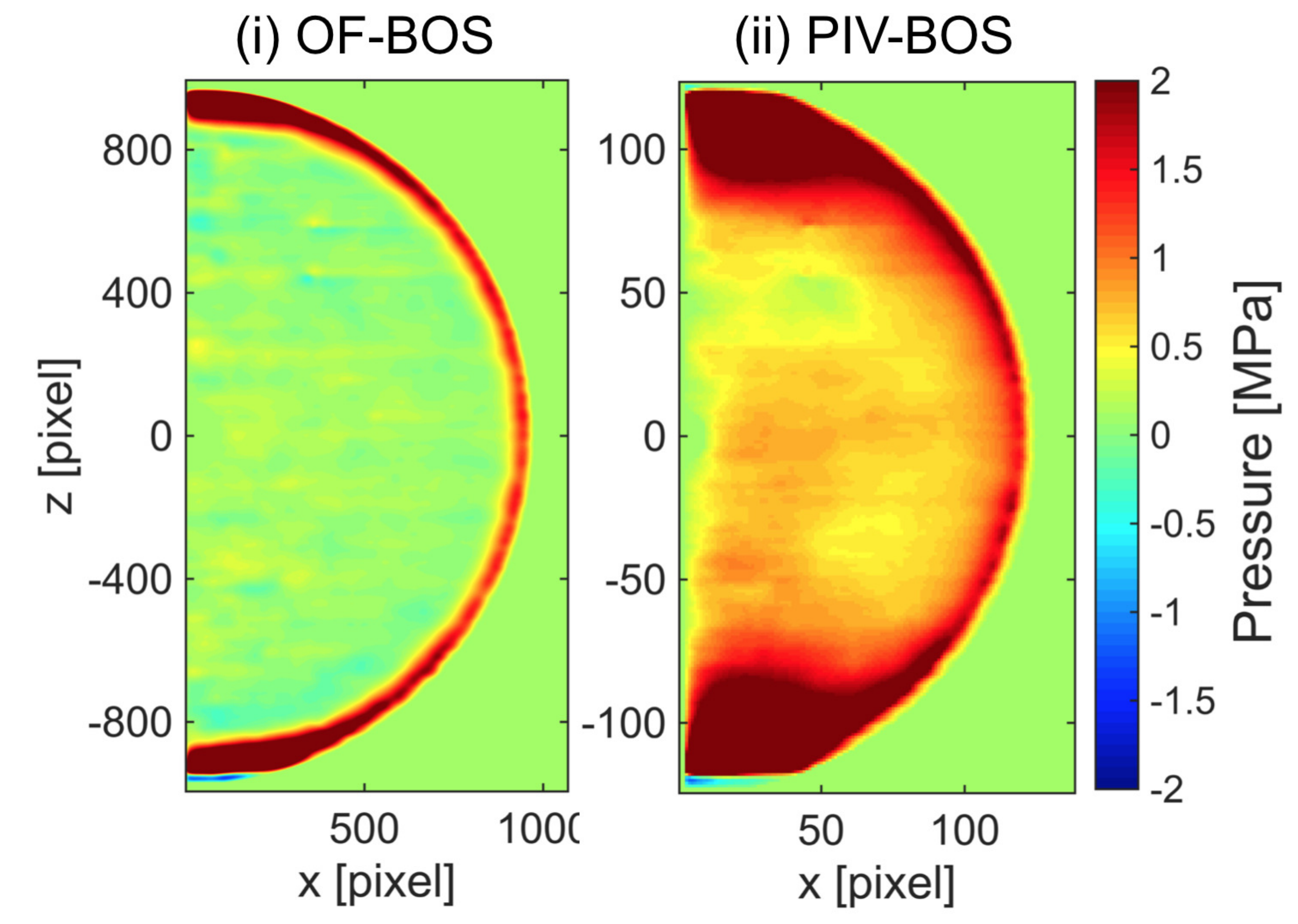}
\caption{The reconstructed pressure fields obtained by using (i) the OF-BOS and (ii) the PIV-BOS.}
\label{Fig10}       
\end{figure}

\begin{figure}[b]
\includegraphics[width=0.8\textwidth]{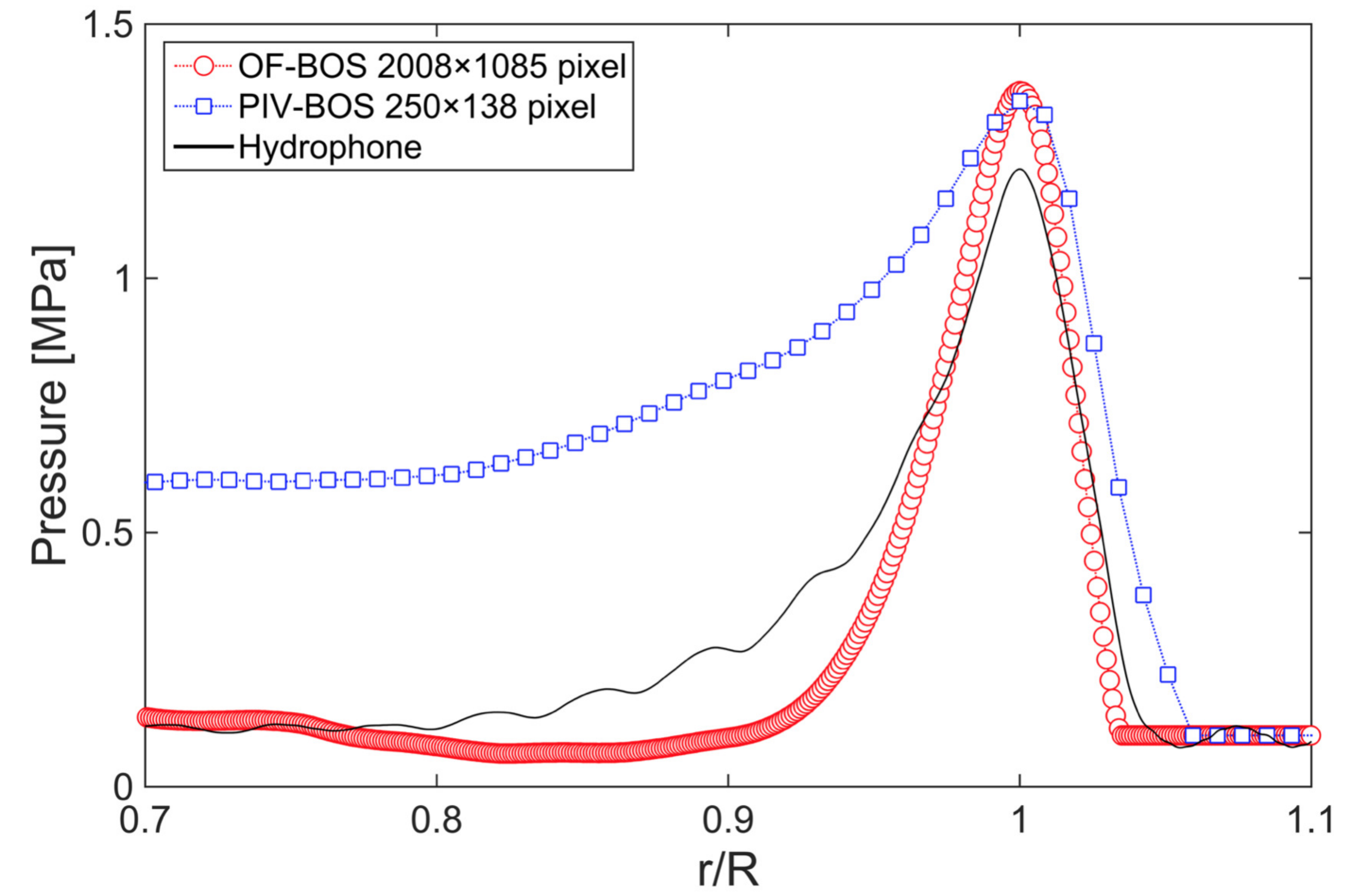}
\caption{Comparisons between the pressure profiles obtained by using the OF-BOS, the PIV-BOS and the hydrophone, where $R$ is the radius of the shock wave.}
\label{Fig11}       
\end{figure}

\begin{figure*}[h!]
\includegraphics[width=1.00\textwidth]{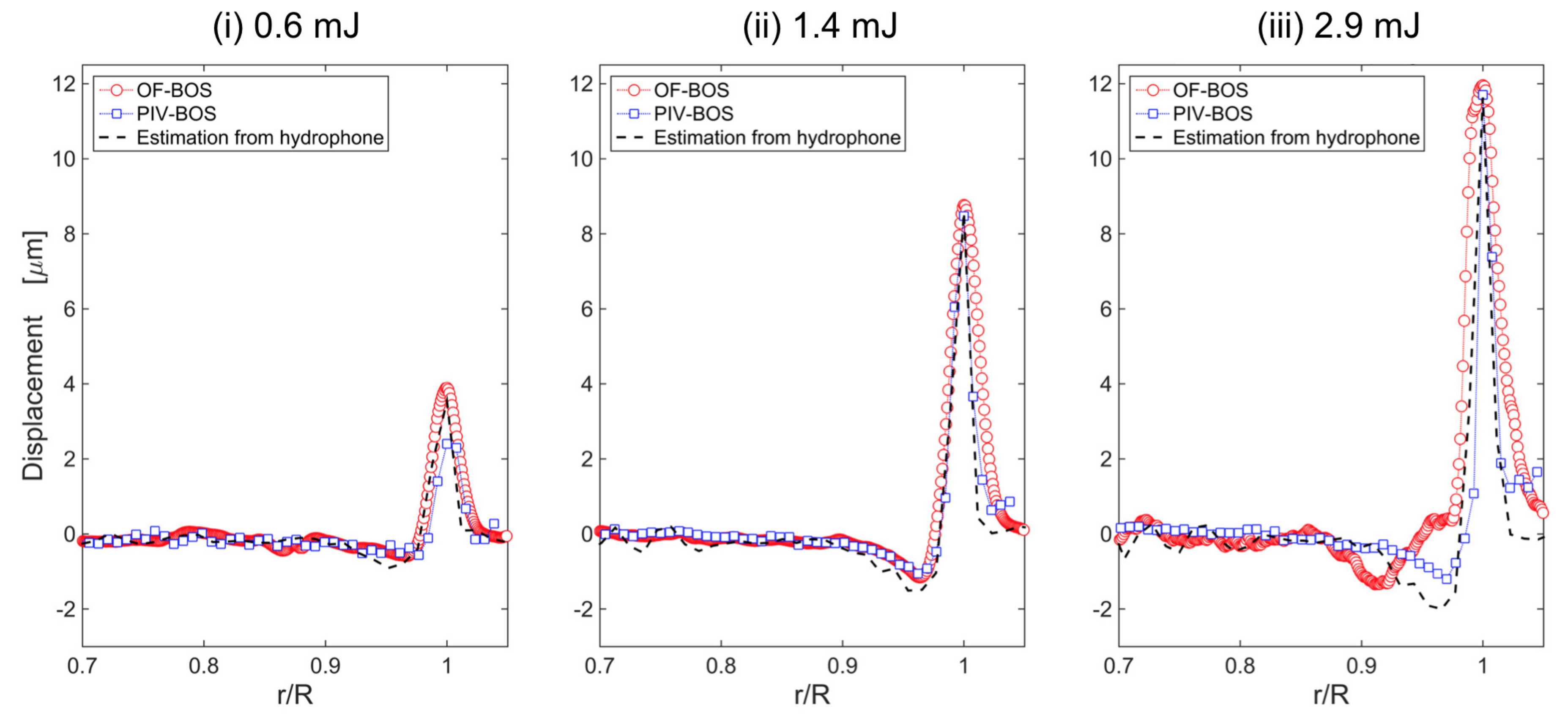}
\caption{The profiles of the measured displacement at three levels of the laser energy, where the OF-BOS and the PIV-BOS data are obtained along the ray aligned with the hydrophone.}
\label{Fig12}       
\end{figure*}

\begin{figure*}[h!]
\includegraphics[width=1.00\textwidth]{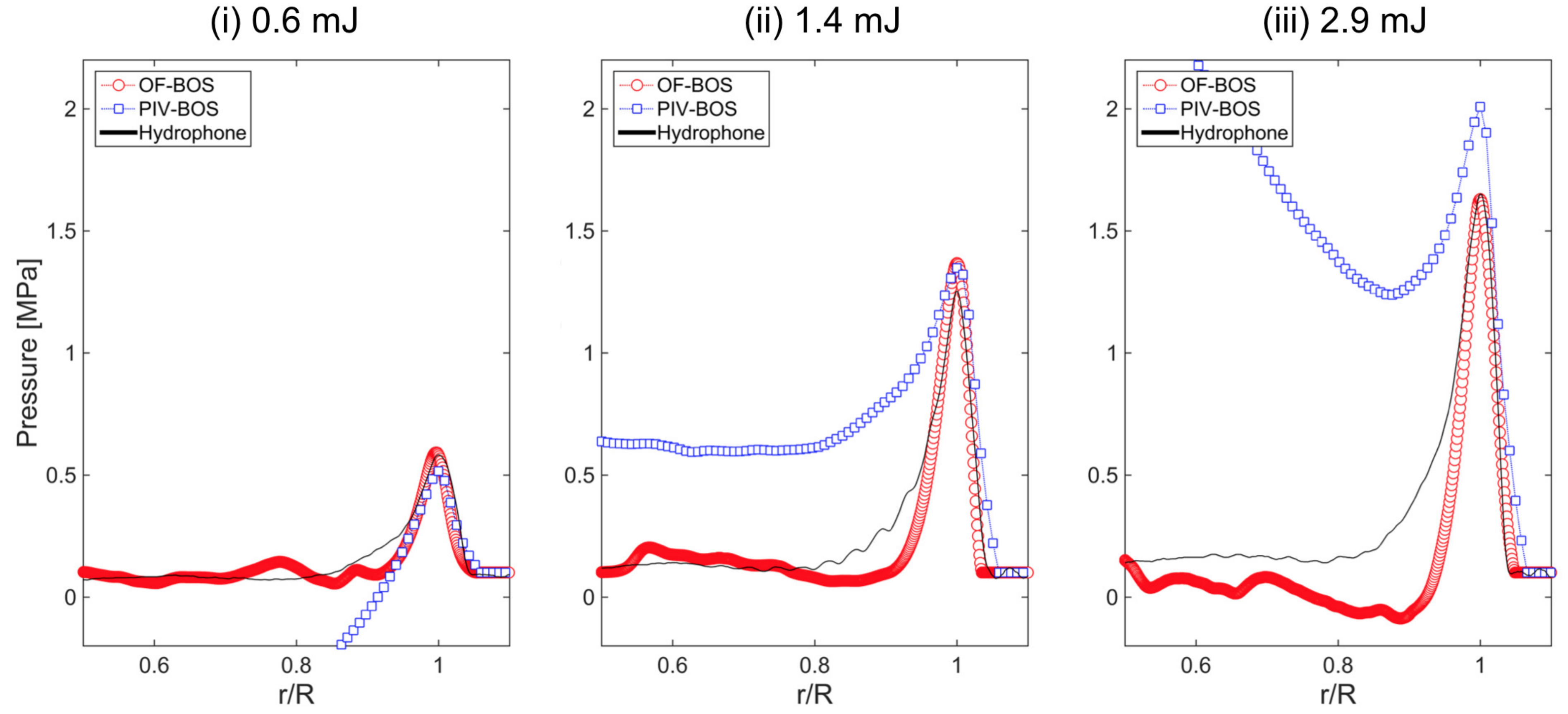}
\caption{Typical pressure distributions of the laser-induced shock wave at three levels of the laser energy, where the OF-BOS and the PIV-BOS data are obtained along the ray aligned with the hydrophone and R is the radius of the shock wave.}
\label{Fig13}       
\end{figure*}

\begin{figure}[h]
\begin{center}
\includegraphics[width=0.50\textwidth]{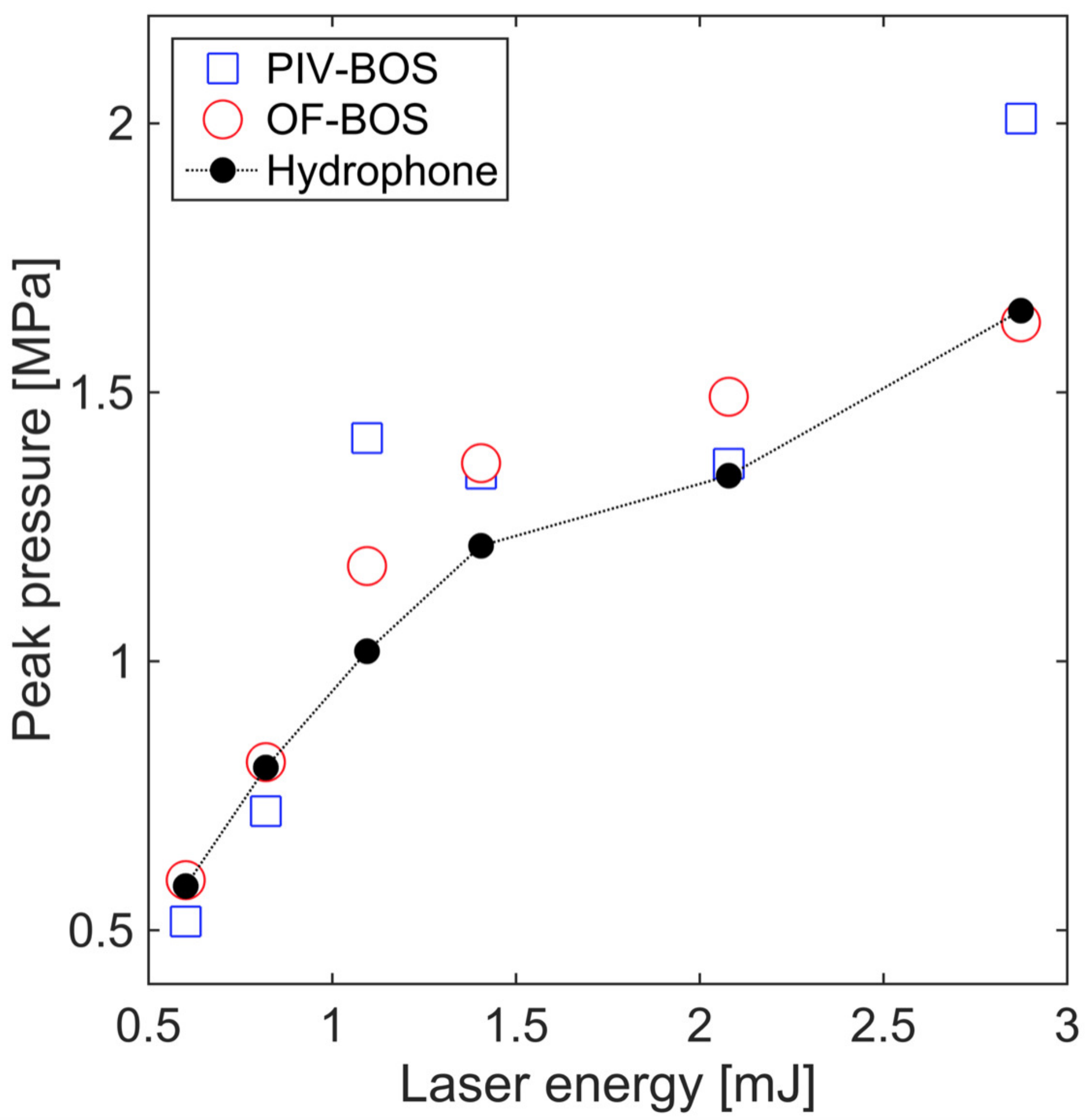}
\caption{The peak pressure value of the shock wave as a function of the laser energy, where the OF-BOS and the PIV-BOS data are obtained along the ray aligned with the hydrophone.}
\label{Fig14} 
\end{center}      
\end{figure}

A typical case with the laser power of 1.2 mJ is considered to compare the results obtained by using the OF-BOS and the PIV-BOS.  
Figure \ref{Fig8} shows the displacement magnitude fields obtained by using the OF-BOS and the PIV-BOS.  
The OF-BOS gives 2008$\times$1085 vectors while the PIV-BOS gives 250$\times$138 vectors in the same domain in Fig. \ref{Fig8}.  
In optical flow computations, the Lagrange multipliers for the Horn-Schunck estimator and the Liu-Shen estimator are 20 and 2000, respectively.  
In PIV computations, the window size is iteratively reduced from the initial size of 32$\times$32 pixels to 16$\times$16 pixels and then to 8$\times$8 pixels.  
Both the OF-BOS and the PIV-BOS capture the sharp change induced by the shock wave in the displacement fields.  
As expected, the cross-correlation computations in windows in the PIV-BOS tend to smooth out sharp features like the shock waves.  
Figure $\ref{Fig9}$ shows the profiles of the measured displacement along a ray aligned with the hydrophone, where the averaging operation over 50 lines for the OF-BOS and 20 lines for the PIV-BOS in the $z$-direction is applied to reduce the noise.  
The displacements were estimated from pressure data by using Eqs. (3)-(6) with an assumption of constant speed of propagation. 
It is indicated that both the OF-BOS and the PIV-BOS give results consistent with that given by the hydrophone at that location.  
In particular, the displacement induced by the shock wave is well captured by both the techniques.  
Nevertheless, the OF-BOS achieves a much higher spatial resolution.  

Figure \ref{Fig10} shows the reconstructed pressure fields obtained by using the OF-BOS and the PIV-BOS.  
The OF-BOS capture correctly the sharp pressure change across the shock wave, while the PIV-BOS has a larger error in the reconstructed pressure field particularly near the `north pole' and `south pole' due to its low spatial resolution.  
For quantitative comparison, as shown in Fig. \ref{Fig11}, the pressure profiles reconstructed based on the displacement vector fields given by the OF-BOS (2008$\times$1085 vectors) and PIV-BOS (250$\times$138 vectors) are plotted against the data obtained by the hydrophone.  
The pressure profile given by the OF-BOS along a ray aligned with the hydrophone is consistent with that given by the hydrophone.  
In contrast, the pressure profile given by the PIV-BOS exhibits a considerably broader distribution extending to the inner region although its peak value at that location agrees with the data given by the hydrophone.  
Clearly, the OF-BOS provides the more accurate reconstruction of the pressure field of the shock wave.  
In contrast, the PIV-BOS has a much larger deviation from the data given by the hydrophone, which is particularly contributed by the large errors near the `north pole' and `south pole' (as shown in Fig. \ref{Fig10}).  
In this case, the major issue of the PIV-BOS is its much lower spatial resolution that tends to corrupt the tomographic reconstruction of the shock wave.


\subsection{Effect of laser energy}
\label{sec:Effect of laser energy}

Measurements at different levels of the laser power were conducted to further compare the data obtained by using the OF-BOS, the PIV-BOS and the hydrophone.  
Figure \ref{Fig12} shows the typical displacement distributions induced by the laser-induced shock wave at three levels of the laser energy, where the OF-BOS and PIV-BOS data are obtained along the ray aligned with the hydrophone.  
Overall, the OF-BOS and PIV-BOS data are in good agreement with the hydrophone data in the low and medium levels of the laser energy (0.6 and 1.4 mJ).  
In the cases of the higher levels of the laser energy (2.1 and 2.9 mJ) with larger displacements, a coarse-to-fine iterative scheme is adopted to improve the accuracy of optical flow computation (\citet[]{liu2012flow, liu2015comp}).  
In this scheme, images are initially downsampled by 2 for a coarse-grained velocity field and then a refined velocity field with the full image resolution is obtained in iterations for correction of the large displacements.  
Two and three iterations are applied to the cases of 2.1 and 2.9 mJ, respectively.  
Figure \ref{Fig13} shows the corresponding pressure distributions of the laser-induced shock wave at three levels of the laser energy.  
The OF-BOS is able to detect the weak shock wave and the pressure peaks of the shock wave are consistent with those given by the hydrophone.  
Figure \ref{Fig14} shows the peak pressure value of the shock wave as a function of the laser energy, indicating the favorable comparison between the data obtained by using the OF-BOS and the hydrophone.  
The PIV-BOS also detects the peak pressure in the lower energy cases, but the pressure distribution inside the shock front does not match that of hydrophone.


\subsection{Effect of spatial resolution}
\label{sec:Effect of spatial resolution}

\begin{figure*}[h!!]
\begin{center}
\includegraphics[width=1\textwidth]{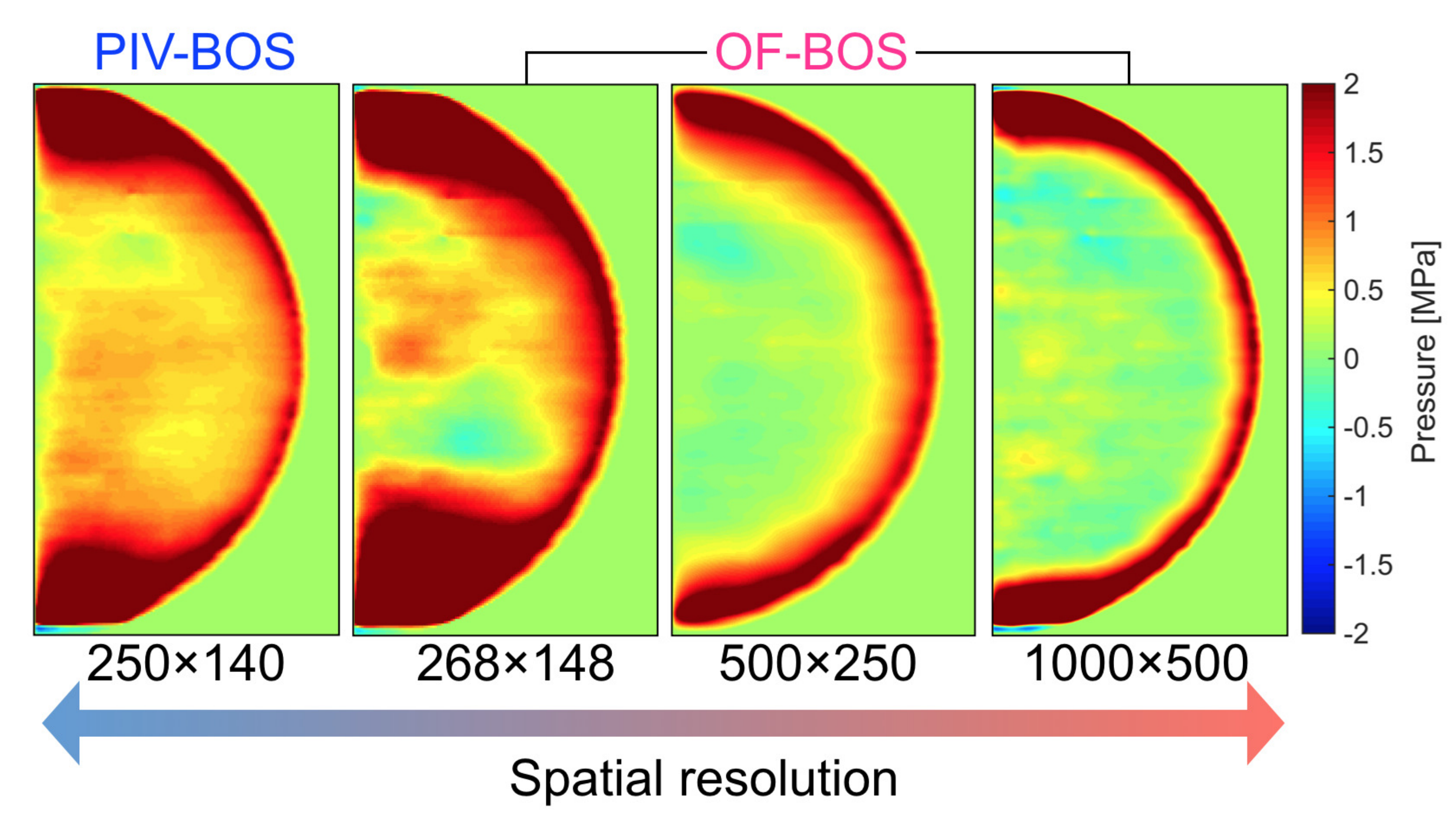}
\caption{The reconstructed pressure fields obtained by the OF-BOS based on the selectively downsampled displacement fields to show the effect of the spatial resolution on the tomographic reconstruction, where the field obtained by the PIV-BOS is shown for reference.}
\label{Fig15}
\end{center}       
\end{figure*}

\begin{figure*}[h!!]
\begin{center}
\includegraphics[width=1\textwidth]{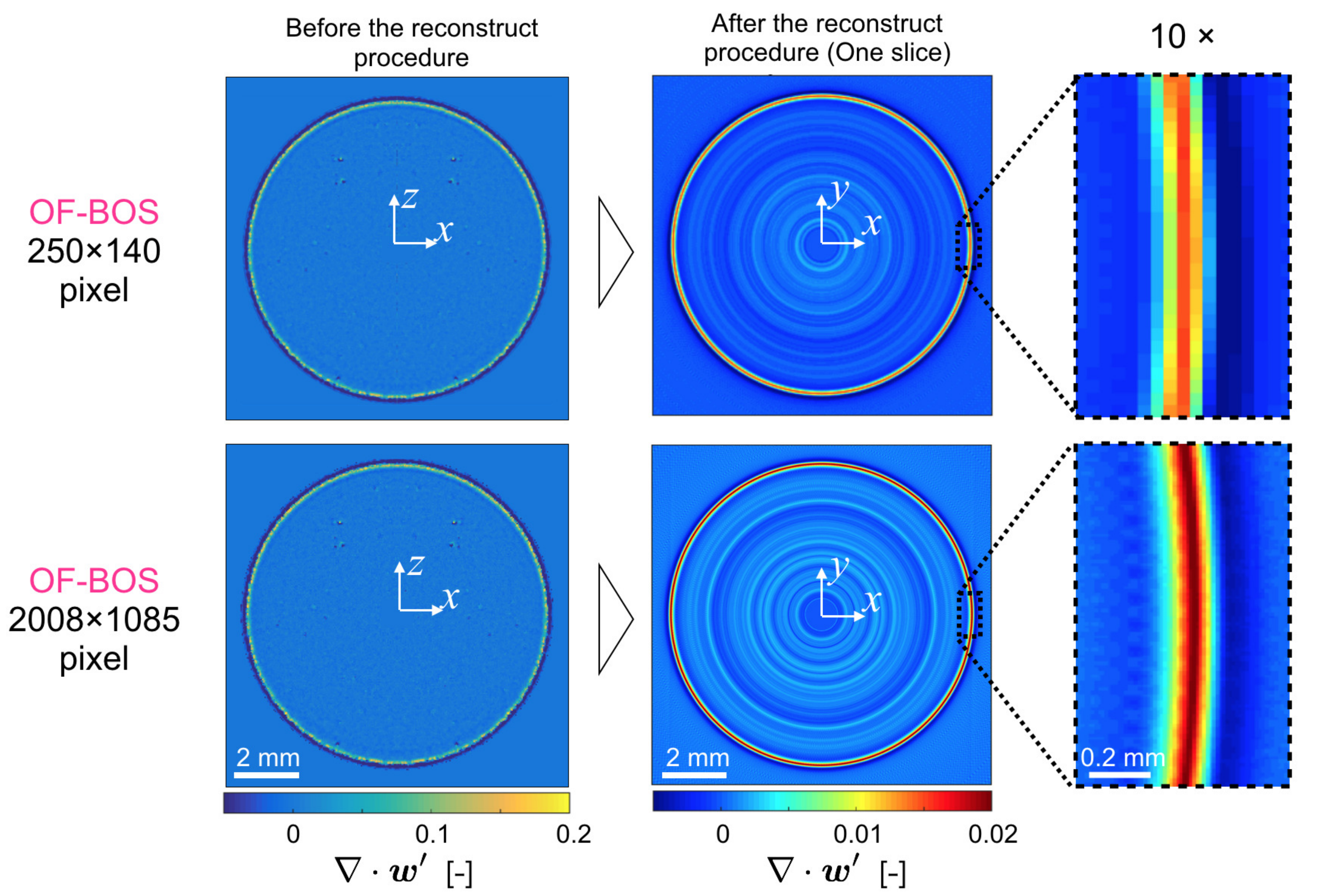}
\caption{The effect of the spatial resolution of the field of the divergence of the displacement vector on the tomographic reconstruction on the pressure field of the shock wave.}
\label{Fig16}
\end{center}       
\end{figure*}

\begin{figure*}[h!!]
\begin{center}
\includegraphics[width=1\textwidth]{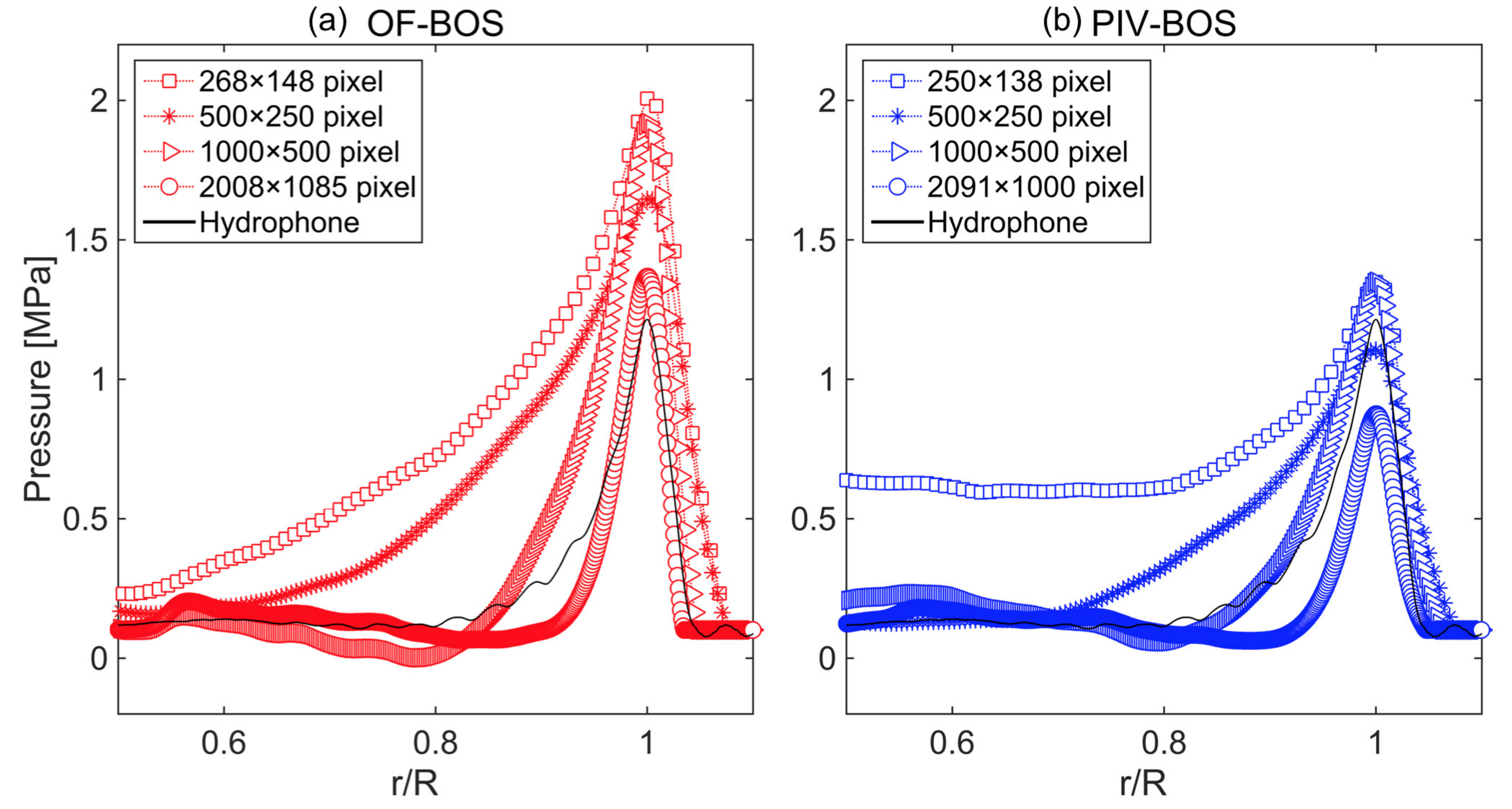}
\caption{The effect of the spatial resolution of the displacement field on the pressure profiles along the ray aligned with the hydrophone: (a) the OF-BOS, and (b) the PIV-BOS, where $R$ is the radius of the shock wave.}
\label{Fig17}    
\end{center}   
\end{figure*}

It is conjectured that the spatial resolution of the extracted displacement field would have a significant impact on the tomographic reconstruction of the pressure field of the shock wave.  
This is because the spatial derivatives of the displacement vector field should be calculated accurately for the tomographic reconstruction of the divergence of $\nabla \cdot {\mbox{\boldmath $w$}}^{\prime}$, which will have the direct influences on the extracted density and pressure fields.  
To examine this problem, the displacement field of 2008$\times$1085 vectors in the half-circle domain obtained by the OF-BOS is downsampled into several fields with the lower resolutions.  
Then, the tomographic reconstruction procedures are applied to these fields to observe the effect of the spatial resolution.  
Figure \ref{Fig15} shows the reconstructed pressure fields based on the selectively downsampled displacement fields, where the field obtained by the PIV-BOS is also shown for reference.  
It can be observed that the quality of the reconstructed pressure field is degraded as the spatial resolution decreases.  
The degraded result given by the PIV-BOS is particularly obvious.  

The quality of the pressure field particularly near the shock wave directly depends on the spatial resolution of the reconstructed field of the divergence of the displacement field, which can be illustrated in Fig. \ref{Fig16}.  
Furthermore, as shown in Fig. \ref{Fig17}, the effect of the spatial resolution on the reconstructed pressure field can be quantitatively observed in the pressure profiles along the ray aligned with the hydrophone.  
For the OF-BOS, the displacement field of 2008$\times$1085 vectors is downsampled to that of 268$\times$148 vectors. 
As shown in Fig. \ref{Fig17}(a) the reconstructed pressure distribution exhibits that the pressure peak is broaden increasingly with the decrease of the spatial resolution and the peak value deviates from the data given by the hydrophone.  
In addition, the displacement field obtained by the PIV-BOS can be interpolated to generate a pseudo high resolution field of 2091$\times$1000 vectors.  
As shown in Fig. \ref{Fig17}(b), such interpolation could indeed improve the quality of the reconstructed pressure field even though the peak value becomes lower.


\section{Conclusion}
\label{Conclusion}

In BOS measurements of a shock wave induced by a laser in water, it is critical to obtain the displacement vector field with the high spatial resolution.  
In this work, the physics-based optical flow method is incorporated into the BOS technique, simply called the OF-BOS, which can obtain the displacement vector field from BOS images at the theoretical resolution of one vector per pixel.  
The tomographic reconstruction for a spherical laser-induced shock wave is proposed, which is applied to the fields of the projected divergence of the displacement vector at cross sections in the vertical direction.  
Superposition of the tomographic solutions at these cross sections yields the local field of the divergence of the displacement vector in the domain of the spherical shock wave.  
Then, the local density field is obtained by solving the Poisson's equation and further the corresponding pressure field is calculated.  

The OF-BOS gives the pressure field consistent with the result obtained by the hydrophone.  
In contrast, the BOS with the conventional cross-correlation method in PIV (simply called the PIV-BOS) fails to correctly reveal the shock wave in the reconstructed pressure field due to its much lower spatial resolution.  
It is found that the quality of the tomographic reconstruction of the divergence of the displacement vector field significantly depends on the spatial resolution of the extracted displacement vector field. 
Therefore, the reconstructed density and pressure fields of the spherical shock wave are directly affected by the spatial resolution.  
Clearly, the OF-BOS has the advantages over the PIV-BOS in this aspect.

\begin{acknowledgements}
This work was supported by JSPS KAKENHI Grant Number 26709007 from the Japan Society for the Promotion of Science.  This work was financially supported by Institute of Global Innovation Research at Tokyo University of Agriculture and Technology.
\end{acknowledgements}

\bibliographystyle{spbasic}      
\bibliography{Reference}   

\begin{thebibliography}{21}
\providecommand{\natexlab}[1]{#1}
\providecommand{\url}[1]{{#1}}
\providecommand{\urlprefix}{URL }
\expandafter\ifx\csname urlstyle\endcsname\relax
  \providecommand{\doi}[1]{DOI~\discretionary{}{}{}#1}\else
  \providecommand{\doi}{DOI~\discretionary{}{}{}\begingroup
  \urlstyle{rm}\Url}\fi
\providecommand{\eprint}[2][]{\url{#2}}

\bibitem[{Atcheson et~al(2009)Atcheson, Heidrich, and
  Ihrke}]{atcheson2009evaluation}
Atcheson B, Heidrich W, Ihrke I (2009) An evaluation of optical flow algorithms
  for background oriented schlieren imaging. Exp Fluids 46(3):467--476

\bibitem[{Brujan(2010)}]{brujan2010cavitation}
Brujan E (2010) Cavitation in Non-Newtonian Fluids: With Biomedical and
  Bioengineering Applications. Springer Science \& Business Media

\bibitem[{Feng et~al(2002)Feng, Okamoto, Tsuru, Madarame, and
  Fumizawa}]{feng2002visualization}
Feng J, Okamoto K, Tsuru D, Madarame H, Fumizawa M (2002) Visualization of {3D}
  gas density distribution using optical tomography. Chem Eng J 86(3):243--250

\bibitem[{Hayasaka et~al(2016)Hayasaka, Tagawa, Liu, and
  Kameda}]{hayasakameasurement}
Hayasaka K, Tagawa Y, Liu T, Kameda M (2016) Measurement of a laser-induced
  underwater shock wave by the optical-flow-based background-oriented schlieren
  technique. In: Proceedings of the 18th International Symposium on Application
  of Laser and Imaging Techniques to Fluid Mechanics, Lisbon, Portugal, pp
  249--257

\bibitem[{Horn and Schunck(1981)}]{horn1981determining}
Horn BK, Schunck BG (1981) Determining optical flow. Artificial intelligence
  17(1-3):185--203

\bibitem[{Klaseboer et~al(2007)Klaseboer, Fong, Turangan, Khoo, Szeri, Calvisi,
  Sankin, and Zhong}]{klaseboer2007interaction}
Klaseboer E, Fong SW, Turangan CK, Khoo BC, Szeri AJ, Calvisi ML, Sankin GN,
  Zhong P (2007) Interaction of lithotripter shockwaves with single inertial
  cavitation bubbles. J Fluid Mech 593:33--56

\bibitem[{Liu and Shen(2008)}]{liu2008fluid}
Liu T, Shen L (2008) Fluid flow and optical flow. J Fluid Mech 614:253--291

\bibitem[{Liu et~al(2012)Liu, Wang, and Choi}]{liu2012flow}
Liu T, Wang B, Choi DS (2012) Flow structures of {J}upiter's {G}reat {R}ed
  {S}pot extracted by using optical flow method. Phys Fluids 24(9):096,601

\bibitem[{Liu et~al(2015)Liu, Merat, Makhmalbaf, Fajardo, and
  Merati}]{liu2015comp}
Liu T, Merat A, Makhmalbaf M, Fajardo C, Merati P (2015) Comparison between
  optical flow and cross-correlation methods for extraction of velocity fields
  from particle images. Exp Fluids 56(8):1--23

\bibitem[{Meier(2002)}]{meier2002computerized}
Meier G (2002) Computerized background-oriented schlieren. Exp Fluids
  33(1):181--187

\bibitem[{Merzkirch(2012)}]{merzkirch2012flow}
Merzkirch W (2012) Flow visualization. Elsevier

\bibitem[{Murphy and Adrian(2011)}]{murphy2011piv}
Murphy MJ, Adrian RJ (2011) {PIV} through moving shocks with refracting
  curvature. Exp Fluids 50(4):847--862

\bibitem[{Raffel(2015)}]{raffel2015background}
Raffel M (2015) Background-oriented schlieren {(BOS)} techniques. Exp Fluids
  56(3):1--17

\bibitem[{Raffel et~al(2000)Raffel, Richard, and
  Meier}]{raffel2000applicability}
Raffel M, Richard H, Meier G (2000) On the applicability of background oriented
  optical tomography for large scale aerodynamic investigations. Exp Fluids
  28(5):477--481

\bibitem[{Tagawa et~al(2012)Tagawa, Oudalov, Visser, Peters, van~der Meer, Sun,
  Prosperetti, and Lohse}]{tagawa2012highly}
Tagawa Y, Oudalov N, Visser CW, Peters IR, van~der Meer D, Sun C, Prosperetti
  A, Lohse D (2012) Highly focused supersonic microjets. Phys Rev X
  2(3):031,002

\bibitem[{Tagawa et~al(2013)Tagawa, Oudalov, El~Ghalbzouri, Sun, and
  Lohse}]{tagawa2013needle}
Tagawa Y, Oudalov N, El~Ghalbzouri A, Sun C, Lohse D (2013) Needle-free
  injection into skin and soft matter with highly focused microjets. Lab Chip
  13(7):1357--1363

\bibitem[{Thielicke and Stamhuis(2014)}]{thielicke2014pivlab}
Thielicke W, Stamhuis E (2014) {PIV}lab--towards user-friendly, affordable and
  accurate digital particle image velocimetry in {MATLAB}. J Open Res Software
  2(1)

\bibitem[{{van}~Hinsberg and R{\"o}sgen(2014)}]{van2014density}
{van}~Hinsberg N, R{\"o}sgen T (2014) Density measurements using near-field
  background-oriented schlieren. Exp Fluids 55(4):1--11

\bibitem[{Venkatakrishnan and Meier(2004)}]{venkatakrishnan2004density}
Venkatakrishnan L, Meier G (2004) Density measurements using the background
  oriented schlieren technique. Exp Fluids 37(2):237--247

\bibitem[{Wang et~al(2015)Wang, Cai, Shen, and Liu}]{wang2015analysis}
Wang B, Cai Z, Shen L, Liu T (2015) An analysis of physics-based optical flow.
  J Computational and Applied Mathematics 276:62--80

\bibitem[{Yamamoto et~al(2015)Yamamoto, Tagawa, and
  Kameda}]{yamamoto2015application}
Yamamoto S, Tagawa Y, Kameda M (2015) Application of background-oriented
  schlieren {(BOS)} technique to a laser-induced underwater shock wave. Exp
  Fluids 56(5):1--7

\end{thebibliography}


%
%

\end{document}